# Hybrid Analog Signal-Based Models of Computation

T. E. Raptis[1,2,3]

1. Department of Informatics and Telecommunications, University of Peloponnese, Tripoli, Greece, 22131;
2. National Center for Science and Research "Demokritos", Division of Applied Technologies, Athens, Greece, 15341;
3. Physical Chemistry Lab., Chemistry Department, National Kapodistrian University of Athens, Athens, Greece, 15784;
* Correspondence: raptis.theophanes@gmail.com

*Abstract:* The present work attempts both a review of previous methods for transferring digital and symbolic computations in an analog or optical substrate and also to offer certain alternatives not yet fully explored. The essential difference from previous cases lies in the merging of general signal processing and computational theory with some emphasis on the foundations of computations and its logico-mathematical background and possible connections with fundamental physical processes. The technologies proposed could among other things be used to turn a standard RF network into a complete, autonomous holographic or distributed computational medium. Some applications for Multi-Agent Systems are also examined near the end.

## 1. Introduction

The effort to find alternatives to the prevailed standard is as old as the development of the digital computer itself. The first digital computer to appear almost immediately after the seminal paper by Alan Turing on his universal machine (UTM) [1] came with Konrad Zuse's programmable Z4 machine [2] based on relays as early as 1941 in Berlin with further development ceasing due to WWII at 1945. It is only ten years after that Von Neumann submits a patent to be granted at 1957, claiming an entirely new, nonlinear mechanism for storing binary information [3], [4]. At about the same time, realization of a similar device also takes place in Japan by Eiichi Goto with his nonlinear "*Parametron*" circuit [5]. Later versions based on superconducting Josephson junctions were known at the 1960s as the "*Quantum Flux Parametron*" which allowed great speed gains at the cost of a deep cooling system. It was superseded at the 1990s by the "*Rapid Single Flux Quantum*" giving rise to the RSFQ logic in contrast with TTL or CMOS logic [6] which is still considered as a potential disruptive technology nowadays [7].

These early attempts can be brought into the same general class of oscillator-based computing. Recently there was a resurgence of interest in this paradigm with work by Izhikevich [8] on neural oscillators as well as Roychowdhury in his "*PHLOGON*" model [9], [10]. A more recent comprehensive review with emphasis on certain neural network applications can be found in a recent review by Csaba and Porod [11].

On the other hand, development of purely analog machines operating in continuous time domain and phase space had started with Vannevar Bush's "*Differential Analyzer*"[12],[13],[14] at the 1930s and culminated with the mathematical proof of universality in the seminal Claude Shannon's 1941 work [15] which gave rise to the General Purpose Analog Computer (GPACK). Since then, several other analog models of computation have been proposed including hybrid machines as those of Bauer and West [16], Schmid[17] the Blum-Shub-Smale (BSS) model [18], the R-recursive functions by Moore [19], Computable Analysis(CoAn) [20], the abstract network model of Tucker and Zucker [21] and the developing field of neural networks [22]. Rubel has also proposed an Extended Analog Computer

(EAC) in [23], [24], [25]. Siegelmann was able to prove the capacity of Recurrent Neural Nets (RNN) for Hyper-Turing computation [26] but only in an idealized setting in the absence of noise. An additional biologically inspired proposal for Hyper-computation has been given by Paun [27], [28] with P-Systems, a kind of membrane computing which has not been practically realized as yet.

At a theoretical level, it was soon realized that certain dynamical systems have the potential of both self-organization and distributed computing [29], [30], [31] as well as the ability to emulating digital and symbolic computations [32], [33], [34]. The capacity of certain nonlinear and chaotic maps and dynamical systems to encode the control states of a binary program was effectively shown by direct simulation in a thesis by Sato with the introduction of the class of "*Switching Map Systems*" [35],[36],[37]. This capacity naturally exists in the bifurcation diagrams of such systems over certain critical parameters as in the seminal example of the logistic map [38]. More recently, Bournez *et al.* were able to show the equivalence between the original GPAC and CoAn [39] as well as the ability of the GPAC to effectively emulate any UTM [40]. Some very recent examples include the ARCO compiler for configuration and synthesis of analog devices [41], as well as the JAUNT solver for dynamical system scaling [42], and an actual implementation of a material bifurcating system based on nonlinear structural elements by Raney *et al.*[43]. .

On a different front, attempts towards incorporating digital and symbolic processing in optical computing start in 1970s and 1980s, motivated by the need to speed up computations, simplify circuitry and lower energy consumption. Certain progress in holographic techniques including matched filtering with correlation holography [44], [45] as well as a proposal for a convolution based Fourier processor [46] led to a consideration of all-optical implementations for general purpose computing machines. A series of reviews on subsequent developments can be found in [47], [48], [49], [50].

Since a standard part of computing machinery involves the need for symbol substitution in large symbolic sequences of varying dimensionality, an early effort towards general purpose optical substitution machines starts with work by Murdocca [51], Brenner *et al.*, in [52], [53], Kiamilev *et al.* [54] and others [55], [56], [57]. A different approach based on Fourier processors was later developed by Naughton and Woods known as the Continuous State Machine (CSM) in [58], [59], [60], [61], [62]. One of the latest, important contributions came from the work of Oltean in the Time Delayed Signals (TDS) paradigm for optical computing [63]. While not a schedule for a general purpose machine, the method was proven applicable to a large variety of non-polynomial ("NP-Complete") combinatoric problems including Satisfiability (SAT), Subset-Sum and, Hamiltonian Path problems [64], [65], [66], [67], [68]. A major difficulty appears to be the exponentially increasing energy cost of the signals required to traverse a whole path set of a hidden binary tree structure.

What delineates the major difference between previous and the current attempt to be presented in the next sections is our motivation in using solely linear mechanisms as well as linear time-invariant (LTI) filters to design a general purpose machine in terms of signal processing methods. Moreover, we aim at solving wherever possible the dissipation problem, bypassing the need for "read/write" operations via a special conservative protocol for analog signal synthesis based on permutations of frequency combs.

In a previous work [69], the author proposed an alternative signal based implementation of simple universal machines known as Cellular Automata (CA) in any dimension. It was also shown that a reversible realization is possible with an exchange between reading and writing operations with permutations. Later on, a different but strongly related example of pumping computations out of a noisy substrate was given in [70] based on certain equivalence between UTMs and a kind of asynchronous CA. The purpose of the present work is to expand the original theme of the Fourier encodings used in [69] and show the possibility of more general sequential and/or functional machines as a combination of waveform synthesizers and convolution filters.

In the next section we review some important notions from the foundations of modern programming languages and we introduce the protocol required for realizing a basic processor as a kind of waveform synthesis device. We also proceed with a low level, special RISC type of architecture as an example. In section III, we examine the possibility of direct implementation of a rewriting grammar which can serve as the basis for special kinds of substitution and functional languages and similar abstract machines. In section IV, we also consider a possible practical application in the case of a recently introduced CA-based machine learning algorithm and its possible transfer into a distributed computing paradigm appropriate also for future robotic swarms where computation can be merged with communication channels. We close with a more speculative discussion on the consequences of such wireless distributed computational models.

## 2. Signal-based computers

*2.1 Syntax and dynamics: Abstract Definitions*

Most current implementations of computing machinery are based on either a direct translation of code into a sequential combinatory logic via logical gates under the general term of "Digital Design", or in some "functional" form of approximation via a limiting "learning" procedure. Alternatively, the latter can also be described via a tuple of characteristic functions or Boolean indicators capable of uniquely describing a particular subset of solutions to a problem such that the total subset can be isolated via a product of such indicators or via an equivalent set theoretic calculus. While very general, the last approach suffers from the difficulty of certain hard combinatoric and other problems where a sufficient set of indicators cannot be directly expressed in analytical form known in advance and certain approximations via sequential search or "sieve" processes have to be used.

Regarding computational architectures, the prevailing paradigm remains that of the standard invented by Eckert and Von Neumann [71], more than 80 years ago while its roots can be traced as far back as the first design of the "Analytic Engine" by Charles Babbage. Most important evolutions concern the development of programming languages in the class of procedural or "imperative" programming which can often be read as abstract machine descriptions only to be emulated by standard combinatory circuits. The most notable evolution concerns the appearance of structured programming with the two seminal papers by Böhm in which he posits the fundamental Structured Program Theorem (SPT) [72],[73], [74] as well as the appearance of so called Reduced Instruction Set Computers (RISC) of which we shall use a special subset known as Ultimate RISC or One Instruction Set Computer (OISC) as an example later on. A very early abstract implementation of URISC machines had been prescribed by Wang in his W-Machine [75], a simple 4-instruciton variant of a UTM. Ater, in a pair of seminal papers Lambel [76] and Melzak [77], further simplified such a construct with their infinite "Abacus". Notably, one such low cost, mini supercomputer realization was recently constructed by Mazonka and Kolodin [78]

From a purely theoretical standpoint, the algebraic theory of finite automata [79] characterizes all such language structures via a particular mapping between a Syntactic Monoid to a Semantic Monoid which is what allows "interpreting" a program seen as a collection of symbolic strings. In mathematical parlor, it is often said that a language is "*recognizable*" by a certain automaton if such a mapping exists. In order to also incorporate general recursion and iteration one also needs to equip a syntactic monoid with the structure of a left or right associative (ordered) semi-group. Furthermore, a syntactic monoid can always be encoded into some form of Gödelian encoding, a fact that we shall use later on in order to be able to transfer any such into an appropriate signal modulation scheme. We shall heretofore refer to this strategy as an "*Arithmetization*". A toolbox for analyzing generic small automata based on the natural self-similarity of a hierarchy of Hamming spaces has been briefly introduced by the author in [70].

Notably, there is a way to trivialize the additional need for an interpreter due to the fact that there are ways to transfer such language structures into simpler substitution automata, also known as "Rewriters" [80], [81] that have the same universality class. In such a way, the need for additional semantics can be transferred into some Look-Up Table (LUT) to be consulted only at the beginning and the end of a computation. This will also leads us away from imperative and closer to functional programming concepts in later sections.

A generic characteristic of all computing machinery that will be helpful for moving away from standard, digital design starts with a minimal abstraction that recognizes any computing activity as a combination of (a) iteration and (b) symbolic substitution or "rewriting" during each step. The latter can be further refined in a (b1) pattern identification via some matched filter mechanism and (b1) some transition mechanism. These elements are common to both the imperative as well as the functional paradigm, the latter originating in early work by Thue [82] on abstract grammars and Church in his calculus of anonymous functions or "lambdas" (λ-calculus) [83], [84], [85] also referred to as first class objects in modern programming. In the original Chruch's calculus, and general functional programming iteration or looping is defined via the so called, "Fixed-Point Combinators"[86], as a kind of recursion performed via functional composition, while in the case of direct implementation on an analog system we must seek an isomorphic physical alternative. In case of cybernetic mechanisms, looping can often be achieved via some feedback mechanism but it may not be sufficiently general for the scope of the present work which goes beyond simple filtering and control.

The simplest example of a direct physical "loop" can be given with the example of two mirrors facing each other thus causing infinite reflections. It is a small step to equip each mirror with some form of curvature and/or some metamaterial structure affecting transmittivity and reflectivity so as to be able to extract a computation as a side effect of successively distorted images. A similar alternative already exists in the form of the so called, "billiard ball" computing paradigm [87], [88] where appropriate choice of both initial and boundary conditions for reflecting a ball's trajectory becomes equivalent to a programmable computing machine.

The previous is suggestive of an immediate generalization where an active element like a transceiver with a passive reflector element becomes capable of iteration. Similarly, we can perform iterations with at least a pair of twin systems each capable of one or more elementary operations before responding to the other thus continually affecting an original message. It is also relatively straightforward to generalize this in the case of Multi-Agent System (MAS) architectures [89], [90]. This is then in contrast with the standard definition of a communication channel where certain error correction methods may be present in order to guarantee the integrity of the received message. Instead, in this interpretation, even the case of distortion/interference could be seen as a type of random or, uncontrollable computation. The essential difference in what follows is then the need for deliberate alterations or distortions of an original message under certain protocols following strict sets of rules.

In what follows we do not address the problem of permanent memory storage. Instead we only assume the capacity of some hybrid system to load a kind of short term memory, analogous to a RAM device, of which the output will then have to be dumped back again to some more permanent storage medium. Such media already exist in the form of certain holographic crystals and other methods which could be easier to adapt to the analog paradigm offered here.

## *2.2 Processing as Waveform Synthesis*

Following [69] we define a temporary memory element via a particular type of a wideband signal modulation. While there are various forms of modulation schemes, the particular type chosen for a binary signal can be given with the aid of what is known as a "Manchester Code" [91] directly in the frequency domain. Specifically, given any binary word of length $L$, we choose to represent it via a bit substitution rule as "0" → "10" and "1" → "01". Assuming then a frequency subdivision of the total signal spectrum into intervals of $2\Delta\omega$, each bit is represented as a 2-permutation or frequency "swap" over a large frequency comb. We will say that the signal $x(t)$ is then a representative of the positive integer $|x|$ via its bit pattern or binary word of maximal length $L$ with zero padding if necessary. The set of all possible such words is often referred to as the *Lth* order "*Hamming Space*" [92] when equipped with the Hamming norm and we shall denote it as $H_L$. There is also a naturally inclusive (self-similar) hierarchy of all such spaces satisfying $H_1 \subset H_2 \subset ..., H_1 = [0,1]$. The frequency comb protocol defined above has then the following desirable properties:

(a) the spectral power remains constant under any possible transitions corresponding to symbol substitutions whenever these transitions are restricted inside the $2^L$ original word dictionary of all possible combinations at least using modular arithmetic,

(b) the spectral content corresponding to each binary symbol can be read without any need for a DFT module via a technique we shall call, a Perturbative Spectral Inference (PSI) and,

(c) any transitions can be performed via the technique of "Additive Synthesis", that is by simply adding and subtracting harmonic signals at prescribed bit-frequency positions.

Notably, since any transition "10" → "01" and vice versa, corresponds to subtracting a harmonic component at say, $n\Delta f$ and adding at $(n + 1)\Delta\omega$ the total is equivalent to adding a single, AM modulated "beat" frequency signal. Before we explicate a method for reading independent bit-frequency components, we should stress the fact that the first property is only possible when very low bitwise operations are to be performed while in a later section we also explore higher alphabet transition variations which allow for more advanced but non-conservative modulation schemes that could be implemented via the communication channels of a large MAS.

In order to apply the PSI technique, we consider an arbitrary state of each bit-frequency component being in either an odd or an even position of the total signal spectrum. Since all components can be in only one of two states, we can use a similar method as in the case of codebooks, that is by adding a pair of harmonic components of prescribed intensities $c_1$ and $c_2$ on two neighboring frequency intervals and using a square detector for the final signal intensity. From Parseval's theorem it is guaranteed that only the squares of the perturbed positions can make a difference. Taking a total power of $Lc_0^2$ where $c_0$ the amplitude of each component allows writing the equations for the perturbed parts as

$$c + c_1^2 + (c_2 + c_0)^2 = c + \delta c_1$$
$$c + (c_1 + c_0)^2 + c_2^2 = c + \delta c_2 \quad (1)$$

In (1), the first part represents the case of the first frequency missing from a pair which is to be interpreted as a "1" while the second corresponds to the first being present corresponding to a "0" while $c$ should equal $(L-1)c_0^2$. The additional constants $\delta c_{1,2}$ can always be chosen so that (1) will give real solutions for the perturbing amplitudes $c_1$ and $c_2$. A simple comparator circuit can then be used directly to infer the exact content of each location.

Additionally, any bitwise NOT complement operation can be directly affected via the introduction of a reference frequency comb of exactly $2Lc_0^2$ total power with all positions filled which we shall denote as $\mathbf{1}(t)$, in which case complementation of any signal x($t$) is immediately given as a signal difference $\overline{x}(t) = \mathbf{1}(t) - x(t)$. Similarly, bitwise shifts corresponding to multiplication by $2^{\pm k}$ for some $k$ inside [1, $L$] can be given by a simple phase factor $\exp(\pm j2\Delta\omega t)$ affecting a shift in the reciprocal domain followed by (a) a band pass filter for cutting frequencies below the allowed interval for the minus sign or (b) addition of the lowest harmonic for the plus sign to represent zeros followed by a low pass filter for modulo arithmetic. Then a whole multiplier circuit would consist of a PSI module successively counting powers of 2 in one of two signals and transforming the other using a succession of shifts.

An optimization problem in this technique concerns the need to find the particular pair member having the lowest sum of digits thus giving the lowest number of shift operations which apparently requires 2$L$ applications of a PSI module on both signals plus an integer in the interval [0, $L$] for the multiplier. A less naïve algorithm can be given from theoretical considerations using an identity for the Hamming norm of two bit patterns via bitwise XOR logic as

$$h(|x|,|y|) \equiv \||x| - |y|\|_1 = S_2(XOR(|x|,|y|)) \quad (2)$$

In (2) we use $S_2$(n) to denote the dist-sum. Using a logical mask for |y| = 0, the all zeros bit pattern allows finding the digit-sum of any integer |x| while |y| = $2^L$ which is the all ones bit pattern gives its complement as $L - S_2$(|x|). We then notice that the truth table of the direct convolution of two signals $x(t)$ and $y(t)$ shown in TABLE I has a structure that results in a total spectral power equaling the complement of the Hamming norm with respect to the total word length $L$ as $(L - \|x(\omega) - y(\omega)\|_1)^2 c_0^2$. This is also the total number of same symbols over which the bitwise XOR would output zero. One can then obtain directly the Hamming distance from the convolution of $x$ and $y$'s complement given via the reference comb as $x(t) \otimes [\mathbf{1}(t) - y(t)]$ and thus the digit-sum can be effectively computed without actual evaluation of the XOR inside (2).

It is then necessary to also find a method for realizing bitwise logical XORing in the particular modulation protocol followed here. On the other hand an antisymmetric combination of the same signals addition and their convolution leads directly to the truth table of XOR but in a modified form which has then to be brought back to the standard modulation protocol. We can then write symbolically

$$XOR(x(t), y(t)) \leftarrow M(x(t) + y(t) - 2x(t) \otimes y(t)) \quad (3)$$

The $M$ symbol stands for a filter responsible to perform the additional transitions "00" → "10" and "11" → "01". Since this only causes in internal permutation and rearrangement of the total signal energy it must be a conservative, "all-pass" adaptive filter. Since this requires some strict dependence on the exact output bit pattern, it represents the only additional complexity due to the need for translating one modulation type to another.

It is in principle possible to bypass this difficulty by the repeated use of an additional matched filter sequentially altering the frequency content via additive synthesis although the trade off in complexity may not be sufficiently smaller. We note in passing that such a problem will only be present in case one needs an emulator of a machine with a CPU-based architecture. It is entirely possible to avoid even that using a direct transcription of functional grammars to signal based machines which we shall examine in later sections.

Theoretically, it is also possible to describe an abstract machine where other whole bitwise operations are possible if an Adder and an XOR filter are present via the relations

$$2AND(|x|,|y|) = |x|+|y|-XOR(|x|,|y|)$$
$$2OR(|x|,|y|) = |x|+|y|+XOR(|x|,|y|)$$
(4)

In the construction of an Adder we shall follow a somewhat different road than the one used in standard digital design which will be the basic element for emulating another type of low level URISC architecture.

TABLE I
BIT-FREQUNCIES MULTIPLICATION/ADDITION

| $X(\omega)$ | $Y(\omega)$ | $X(\omega)Y(\omega)$ | $X(\omega)+Y(\omega)$ |
|---|---|---|---|
| 1 0 | 1 0 | 1 0 | 2 0 |
| 0 1 | 1 0 | 0 0 | 1 1 |
| 1 0 | 0 1 | 0 0 | 1 1 |
| 0 1 | 0 1 | 0 1 | 0 2 |

*2.3 Analog URISC machines*

One of the simplest URISC architectures that has been realized for experimentation can be found under the acronyms of *AddLeq* or *SubLeq* functions which actually mean "Add-and-branch-if-Less-than-or-Equal-to-zero" and "Subtract-and-branch-if-Less-than-or-Equal-to-zero" respectively. These are implemented as ternary operators of the form *ALE*( *a, b, c* ) or *SLE*(*a, b, c*) where all arguments correspond to positive integers serving as pointers to certain memory locations. Their actual functioning is given in terms of some temporary memory locations *m*[*a, b, c*] in which case if $m[a] \pm m[b] > 0$ we set $m[b] = m[a] \pm m[b]$ otherwise a jump is performed at position *m*[*c*] and the process repeats from there. Negative values if met at either *a*, or *b*, can be used for further data input and output respectively.

While we have not explicitly treated addressing in synthesizer machines up to now, we can adopt a simple multiplexing scheme where a large wire bus serves for recirculating all temporary signals to be processed. It is proven [93], [94] that given a particular realization of either the *AddLeq* or *SubLeq* function, these suffice for constructing all other higher level operations of a full assembly based machine by simple functional composition and altering of the parametrization at each step. For a complete emulation by a synthesizer machine, it suffices to provide an efficient Adder or Subtractor implementation as a signal based automaton.

We proceed with the construction of an abstract Unbounded Precision Serial Adder (UPSA) automaton before discussing its explicit signal based implementation. We notice that given the Adder definition and the existence of a NOT complement circuit it is also possible to define a subtraction automaton via the branched function over maximal constant length *L* words

$$Minus(|x|,|y|) = \begin{cases} t(|x|,|y|), & |x| \geq |y| \\ -t(|y|,|x|), & |x| < |y| \end{cases}$$
(5a)

where

$$t(|x|,|y|) = NOT(NOT(|x|)+|y|)$$
(5b)

For the UPSA, we consider first the simple addition of a power of $2^k$ on an arbitrary string |x| of length $L$. This then always follows the below two rules.

(a) If $k \leq L$ and the coefficient |x|(k) in the polynomial representation of base *2* of |x| is zero then we set |x|(k) = 1. If $k > L$ we always set |x|(k) = 1.

(b) If $k \leq L$ and |x|(k) = 1 we set |x|(k) = 0 and |x|(k') = 0 for all subsequent $k' > k$ for which |x|(k') = 1 until we find a *k'*+1 for which |x|(k'+1) = 0 which is then set to 1.

This particular kind of dynamics can then be repeated for all the non-zero bits of the binary expansion of any other word |y| in the summand |x| + |y|. For optimization purposes we consider the polynomial expansions *P*(|x|) and *P*(|y|) of whom the corresponding orders are the sums of digits of each expansion and we choose to successively add the corresponding powers of 2 in the expansion with minimal order using again the method in (2) of the previous section. Then it is possible to extract all non-zero positions of the minimal $S_2$ signal say, |x| in ascending order whereas for any non-zero position *k* of |x| the same position of |y| is scanned with the PSI protocol and if found at a zero state "10", it gets inverted by addition of the appropriate beat signal, otherwise iterative application of complementary beat signal inverts the whole block of subsequent non-zero positions up to the first following zero.

From this and the previous sections we see that most of the fundamental operations required for standard universal computation of any Turing complete imperative language can be performed based solely on power measurements and additive synthesis over a set of appropriate broadband signals with the amount of total spectral energy fluctuating over a constant mean value without significant dissipative effects apart from possible ohmic losses that may call for some minimal amplification stage. There is still an issue regarding bifurcating signals for intermediate power measurements and optimization of additions, subtractions and multiplications since such may temporarily destroy the strict modulation protocol. As these signals cannot be recirculated they must be consumed which could cause entropy production unless a scheme can be found to restore the original modulation in which case total spectral power will remain constant on the average. While this serves as a set of theoretical guidelines for emulating standard existing machines there are reasons to expand this paradigm to both higher alphabets and possibly functional architectures to the degree to which certain primitive syntactic structures prove appropriate for direct implementation in some dynamical systems via Fourier encodings. We examine some of the simplest such rewriting machines in the next section.

## 3. Universal Substitution Automata

The most general class of substitution automata can be traced back in the work of Thue and in the subsequent development of non-reversible, semi-Thue grammars [95]. In a recent work, Suchanek [96] provided an example of a Universal Replacement System (URS) which is based on an analogous substitution automaton which can then be used to incorporate a variety of rules for both, set theory, Churches *λ*-calculus or Shonfinkel's *SK* combinator calculus. Additionally, in a recent report [97], Maier proved the existence of at least one possible emulator of some variant of a quantum computer by a semi-Thue system. We note in passing that previous signal based models of quantum computers have been proposed by Kish [98] and at least a real version was constructed by LaCour [99], [100].

All such substitution automata are characterized by a two column Look-Up Table (LUT) containing pairs of left and right rules with the right part being a replacement for the left part and an initial condition in the form of a random string in some alphabet in base $b \geq 2$. The original Thue grammars concentrated around symmetric, reversible sets of rules while the more general semi-Thue systems may have arbitrary rule lengths. Computation proceeds with the identification of each instance of a substring from the left LUT's column and replacement with the corresponding right column member, the whole process being then termed a "*Reduction*".

Whether replacement will take place deterministically (rightwise or leftwise on subsequent versions of the original string) or stochastically does not matter due to a theorem by Church and Rosser [101], [102] which holds true at least for certain variants of the original Church's calculus. It requires the property of "*Confluency*" associated with rule sets that always have an intermediate term |z| such that for any direct replacement of sub-words as |x| → w|y| there exists a pair of paths |x| → w|z| and |y| → w|z| where *w* any sub-word. Another important property regards the existence of some terminal, canonical or normal forms of strings, towards which any normalizing set of rules converges. These are then equivalent to attractors of ordinary dynamical systems. For arbitrary random sets of rules the question of termination or of the length of a reduction is undecidable in general.

Chaitin originally [103] considered the case of a direct application of a combinator based functional calculus as impractical due to very low speed including the use of so called "*Church's Numerals*" for encoding integers via successors which are close to a unary system. Later, Tromp [104], solved that problem with the introduction of his "Binary Lambda Calculus" (BLC) where actual numerics with any resulting redex expressions can be carried over by additional, dedicated circuitry as the one explained in the previous sections. Despite the fact that search over the simplest possible functional machines is an open question, we may take the BLC paradigm as the most effective existing alternative. We shall not therefore have to deal further with the appropriate set of rules and we shall concentrate on the technical details for a complete transcription of any Thue system into an analog synthesizer machine.

Any such system is reducible to a single iterative loop followed by some form of evaluation of a final expression, the analog of the famous "*eval*" function in Lisp-like languages. We notice that this interpretation is strongly biased towards what is known in computer science as a "*lazy*" evaluation [105]. As already noticed in section II.B, each step requires a pattern matching mechanism followed by a substitution mechanism. In signal processing terminology, the first half can always be satisfied by an appropriate matched filter based on correlation. The second half-step is of higher complexity since the rule sets adopted in the generic case of semi-Thue grammars do not have a constant length. Moreover, the use of the conservative binary modulation protocol appears inappropriate in the case of arbitrary symbolic sequences.

The first step towards efficient realization of such systems requires the full use of advanced modulation schemes used in modern communication systems. These include certain division multiplexing schemes the most common being the Frequency/Wavelength Division Multiplexing (FDM/WDM) [106]. There also exist their orthogonal versions (OFDM/OWDM) [107] which mimic the properties of a Hilbert space by a direct translation to the time domain with the inner product being replaced by a temporal average. This can then be used as an alternative to the PSI method of the previous section for addressing individual symbols. There are by now several more advances in these areas which go beyond the scope of this short report as in the case of spread spectrum methods like carrier interferometry [108], Orthogonal Chirp DMux (OCDM) [109], Wavelet Fractal DMUX [110], [111] and others. For our purposes here it suffices to provide a method for arithmetizing the original Thue construct which could then be adapted to existing or future modulation protocols.

In principle, a direct translation of the iterative substitution in a harmonic encoding would require the combination of a of low-pass and high-pass filtering for separating parts of the signals of differing time lengths which may suffer from certain technical problems in an analog system regarding throughput and other parameters. Instead, it is preferable to translate the original grammar to a different encoding. To this aim we shall use a type of Gödel encoding for bounded strings as mentioned in [70] to express any set of rules. We will also have to expand the previously introduced [69] method of decomposing any iteration into a product of "detect" and "replace" operations where a matched filter was given in the form of a circulant convolution with a filter mask resembling the polynomial representation.

Given a LUT of the form $\alpha_1^i...\alpha_{Li}^i \rightarrow \sigma_1^i...\sigma_{Ri}^i$ with $i = 1,...,N$, the row index and $L_i$, $R_i$ the corresponding lengths, we first isolate the maximal number of unique elements, say $l$ and we set $b = l + 1$, as the basis of the higher alphabet *excluding zero*. For instance, any LUT with only three symbols as *abc → baa, etc.* can be turned in the arithmetic form 123 → 211. This coding can then be used to reduce every left or right word of the original rule set into a single symbol via the correspondence

$$\alpha_1^i + b\alpha_2^i + b^{Li-1}\alpha_{Li}^i \rightarrow n_i \qquad (7)$$

Doing the same for the right part symbols results in relations of the form $n_i \rightarrow m_i$. As a matter of fact, any such rule set could also be given as a list of single large integers $k_i$ and an additional list of lengths for sub-words in which case any LUT is immediately retrievable via the relations

$$\begin{cases} n_i = \mathrm{mod}(k_i, b^{L_i}) \\ m_i = \left\lfloor \dfrac{k_i}{b^{L_i}} \right\rfloor \end{cases} \tag{8}$$

Exclusion of zero serves to drop the strict dependence on $R_i$ without causing confusion for as long as no null words are allowed in the production rules. It also has the consequence that any arithmetized LUT will contain only rule codes that are not ($mod\ b^k$) congruent for any integers $b, k$.

We can now affect any transition using a circulant convolution with the additional, temporary introduction of two "0" symbols at the edges representing cyclic boundaries so as to be able to use a filter mask of the form $c = [b,..., b^{Li-1}, 0...0, 1]$ on any input string $s$. An immediate indicator for any positions corresponding to a left symbol $n_i$ can then be given via a set of roots using the corresponding circulant matrix $C$ to form the set of operators $(\hat{P}_i)\mathbf{s} = \mathbf{C} \cdot \mathbf{s} - n_i \mathbf{1}$.

Alternatively, we may use a list of outputs for all possible neighboring symbols combinations in the total interval $\left[1, b^{Max\{Li\}} - 1\right]$ with the rule set transitions representing the only deviations from a linear graph so as to bring the dynamics in the more immediate form

$$\mathbf{s}_{n+1} \leftarrow D(R(\mathbf{C} \cdot \mathbf{s}_n)) \tag{9}$$

In (9), $R$ stands for the total 1-1 transition rule and $D$ is a necessary decoder/modulator to restore the original lower alphabet since any transitions affect changes in the internal symbol correlations.

Attempts to transfer the $s$ dynamics directly into an OFDM protocol will require several additional technical details that will be part of a forthcoming work. Additionally, as mentioned in section II.B, it is preferable to use at least a transceiver scheme for realizing the necessary intermediate steps in (9). It is sufficient to state here that MAS architectures are known to be in the same universality class with either the Von Neumann architecture or that of Lisp machines. A MAS structure allows utilizing a two step process based on the basic property of circulant filters that admit a diagonal form as $\mathbf{C} = \mathbf{F} \cdot \Lambda \cdot \mathbf{F}^*$ where the diagonal part is defined from the filter mask as $\Lambda_{ii} \leftarrow \tilde{\mathbf{c}}_0 = \mathbf{F} \cdot \mathbf{c}_0$. It is then possible to introduce a distributed ("holographic") representation of an activation field as $\tilde{\mathbf{h}}_n(\omega) = \Lambda \cdot \tilde{\mathbf{s}}_n(\omega)$. This should be followed by a set of $N$ transceivers or, "agents" of a MAS plus a tagging code in $[1,...,N]$ for each rule pair able to demodulate, modify and reemit the new field structure as in (9) via

$$\mathbf{h}_{n+1} \leftarrow \Lambda \cdot \mathbf{F} \cdot (D \circ R)(\mathbf{F}^* \cdot \mathbf{h}_n) \tag{10}$$

For rule sets with the Church-Rosser property, the tagging code could be assigned at random as already stated in the beginning of the section. Apparently, the expression in (10) leads to the possibility of turning a particular communication network into a computational medium. Moreover, this may represent an efficient mechanism for setting up a self-organized swarm of robotic devices or drones. We note in passing, that a previous first attempt for data transmission into multiple agents has been proposed by Hendry [112], [113] based on establishing a set of Zenneck stationary surface modes on corrugated surfaces. We shall next examine the possibility of applying a similar technique to a special case of deep learning algorithm allowing the emergence of a class of immersive, ambient A. I. mechanisms.

4. Reservoir Computing

The last two decades a new paradigm appeared in the field of machine learning and A. I., based on original work on Liquid State Machines (LSM) and Echo State Networks (ESN) by Shomaker [114], Kirby [115], Maas [116], [117], Jaeger [118], [119], collectively known as Reservoir Computing (RC). Jaeger was able to introduce the notion of "*Conceptors*" through which a number of concepts forming a semantic structure can be made to form in the large parametric space of such systems after training.

The original reservoir construct was based on the use of a dynamical system with high dimensionality with prototypes given via Recurrent Neural Nets (RNN). A low dimension vector of inputs are then injected via some regression network as perturbations in a matrix holding many previous instances of this system iterations while, outputs are also pumped via another regression network. The overall dynamics can be expressed as

$$\begin{aligned}\mathbf{x}_{n+1} &= f\left(\mathbf{W}\cdot\mathbf{x}_n + \mathbf{A}\cdot\mathbf{y}_n + \mathbf{B}\cdot\mathbf{z}_n^{In}\right) \\ \mathbf{y}_n &= g\left(\mathbf{C}^{Out}\cdot\left[\mathbf{x}_n;\mathbf{z}_n^{In}\right]\right)\end{aligned} \qquad (11)$$

In (11), "[ ]" represents a concatenation of the input vector with the current reservoir state and **W, A, B, C** are the internal connectivity weight, feedback, input and output matrices respectively while *f* and *g* are standard sigmoid or similar functions.

Since the conditions for learning in this model are fairly general they do not appear to be critically restricted to the use of an RNN as the core, reservoir dynamical system hence, it has been proposed that other alternatives even with a continuous background dynamics could be used [120], [121], [122]. Recently, Yilmaz [123], [124] was able to show that it is possible to use a simple binary reservoir from the collective trajectory of certain binary CA from a class of computationally universal ones as in the case of the rule 110. Similar results were reported in [125], [126], [127], [128].

Using the logic of the previous section and the original work in [69], it is possible to turn a binary CA into a similar, distributed dynamics using the simpler convolution mask $c_0 = [2, ..., 2^{D-1}, 0,...,0,1]$ where $D$ the number of neighboring cells with the activation field this time encoded directly into the frequency domain as $\tilde{\mathbf{h}}_n(\omega) = \mathbf{C}\cdot\tilde{\mathbf{s}}_n(\omega)$ corresponding to the reciprocal signal $\mathbf{h}_n(\tau) = \Lambda\cdot\mathbf{s}_n(\tau)$. The simplified total dynamics of the form $\mathbf{s}_{n+1} \leftarrow R(\mathbf{C}\cdot\mathbf{s}_n)$ can now be turned into an LTI filter cascade by using a root-pole representation of the *R* filter as a map between the increased alphabet of all neighbor states back into a binary alphabet using a rule dependent transfer function as

$$t_R(\omega) = \frac{\prod_n^{2^D - S_2(R)}(h(\omega)-\rho_0)}{\prod_m^{S_2(R)}(h(\omega)-\rho_1)} \qquad (12)$$

In (12), $2^D$ is the total number of possible states for a neighborhood of *D* cells, $S_2(R)$ the one bits in any rule, and $\rho_{0,1}$ correspond to collective cell states where transitions to 0 or 1 should take place. Practically, this allows using a comparator for the output dB level guiding the presence or absence of a particular frequency in a subsequent modulator.

There is an important LTI alternative which allows a direct correspondence with fundamental wave mechanical processes. In the theory of reverberating processes, there is at least one method of a tunable comb filter with adaptable delays known as a "flanging" filter. For a comprehensive review of delay filters the reader may consult Smith's book [129]. A standard delay comb filter is characterized by an infinite number of periodic zeros and poles for forward and backward delays respectively with the minima and maxima of any response function being often called "notches" and "peaks". Using the flanging technique it becomes possible to modulate the delay terms so as to move pitches around a central position. Such a modulation then suffices for realizing an alternative representation of (12) using odd and even positions for 0s and 1s in a sequence of $k\pi$ intervals of the phase response function. A direct physical equivalent of such filters can be given in terms of a cavity with sensors and actuators located on a time varying boundary controlling interference of multiply reflected signals. The sensors can then be used for collecting the flanged signal pass it through the polynomial convolution mask and inject the resulting activation field back into the cavity. More on the significance of these observations are mentioned in the final discussion section.

The direct use of signal based methods for RC opens the way for decentralized, A. I. applications on MAS with mobile units as for instance in the case of self-organized drone swarms able to learn and navigate on certain terrains which is currently lacking. For instance, it is possible to consider a star network formation of agents holding a history of CA instances which then serves as the reservoir. This is to be created as a concatenation $[x_0(t), x_1(t + \tau)..., x_n(t + n\tau)]$. Since each agent will perform an identical filtering to input signals, the result of a bunch of circulating signals will be a large "delay vector" of signals stored in the agent's swarm at any time.

It is further questionable whether the use of certain wave modes could be utilized for a generic RC in a manner similar to later developments in 3D holographic radar technology [130]. Additionally, we notice the existence of at least two breakthrough works on the possibility of "zero power stealth" noisy channels by Kish in [131], as well as noisy computing [132]. This could initiate a whole area of stealth distributed computing if combined with other recent findings on the possibility of vector potential based, "zero-field" communications [133] via the use of the Aharonov-Bohm effect [134], [135]. The possibility of a MAS swarm sharing secured information through at least some short range stealth channels is an intriguing one and deserves separate examination in a future work.

5. Discussion and Conclusions

In the present work the author attempted to provide a comprehensive review of several interconnected topics from computational theory and show their direct correspondence with a class of signal based as well as physical processes. This then served the purpose of overcoming part of the dissipation problem as well as exploring certain versions of the distributed computing paradigm based on blurring the strict limits between communication and computation. The author believes that this opens a new avenue in alternative computing which becomes necessary as we near the end of so called, "Moore's Law". Moreover, since the notion of signal based computing is not necessarily restricted to the frequency domain, further development of similar methods may give rise to a most general class of "field-theoretic" computing with the analogy between the flanging effect and the cavity computer model of last section as an introductory example. We should also stress here the recent controversy that arose out of the work of Kish in the problem of dissipation [136] as well as the actual relationship between physical, thermodynamic entropy and information entropy [137] which begs for a deeper understanding of the connections between computational and physical processes. The author has already proposed the possibility of conceiving noisy backgrounds as a superposition of an at least countable infinity of random computations in [70]. Some interesting possibilities involving unconventional physical channels naturally follow and should be examined in detail in future work.